\documentclass[12pt, amsmath, amssymb, graphicx]{iopart}
\usepackage{iopams}  
\usepackage{physics}
\usepackage{dsfont}
\usepackage{siunitx}
\usepackage{enumitem}
\usepackage{etoolbox}
\usepackage{standalone}
\usepackage{bm}
\usepackage{soul}
\usepackage{comment}

\usepackage[
backend=biber,
style=chem-acs,
sorting=none
]{biblatex}

\begin{document}

\title[]{From Stochastic Hamiltonian to Quantum Simulation: Exploring Memory Effects in Exciton Dynamics}

\author{Federico Gallina$^\dag$, Matteo Bruschi$^\dag$, Barbara Fresch$^{\dag \ddag}$}
\address{\dag Dipartimento di Scienze Chimiche, Università degli Studi di Padova, via Marzolo 1, Padua 35131, Italy}
\address{\ddag Padua Quantum Technologies Research Center, Università degli Studi di Padova, via Gradenigo 6/A, Padua 35131, Italy}

\ead{federico.gallina@unipd.it, barbara.fresch@unipd.it}
\vspace{10pt}
\begin{indented}
\item[]\today
\end{indented}

\begin{abstract}
The unraveling of open quantum system dynamics in terms of stochastic quantum trajectories offers a picture of open system dynamics that consistently considers memory effects stemming from the finite correlation time of environment fluctuations. These fluctuations significantly influence the coherence and energy transport properties of excitonic systems. When their correlation time is comparable to the timescale of the Hamiltonian evolution, it leads to the departure of open system dynamics from the Markovian limit. In this work, we leverage the unraveling of exciton dynamics through stochastic Hamiltonian propagators to design quantum circuits that simulate exciton transport, capturing finite memory effects. In addition to enabling the synthesis of parametrizable quantum circuits, stochastic unitary propagators provide a transparent framework for investigating non-Markovian effects on exciton transport. Our analysis reveals a nuanced relationship between environment correlation time and transport efficiency, identifying a regime of ``memory-assisted" quantum transport where time-correlated fluctuations allow the system to reach higher efficiency. However, this property is not universal and can only be realized in conjunction with specific features of the system Hamiltonian.

\end{abstract}

%
%
%
\maketitle
%
%

\section{Introduction}

The dynamics at the molecular scale results from an intricate balance between coherent quantum evolution and irreversible dissipative processes. Due to the large number of coupled degrees of freedom evolving on overlapping timescales, it is not always obvious how to separate the relevant molecular states that need to be treated explicitly from the remaining part that acts as the surrounding environment, inducing decoherence and dissipation. Still, such a partition is crucial for elucidating phenomena such as charge and energy transfer, which play pivotal roles in photosynthesis, vibrational energy redistribution, and molecular dynamics. In contrast to the dominant paradigm in quantum information, which considers interactions with the environment as detrimental noise hindering the progress towards functional quantum devices, it is well recognized that, in natural systems, the efficiency of excitonic energy transfer may be enhanced when coherent and incoherent contributions cooperate in a regime of Environment-Assisted Quantum Transport (ENAQT) \cite{Rebentrost2009, Zerah-Harush2018, Li2022, Coates2023, Alterman2024}.

The theoretical study of ENAQT requires an effective model of quantum dynamics able to describe different coupling regimes with the environment. The stochastic approach developed by Haken and Strobl \cite{Haken1973} describes the transition from short-time coherent exciton dynamics to long-time incoherent exciton diffusion in a simple and effective framework, thus becoming a reference model for the study of ENAQT \cite{Rebentrost2009, Wu2012, Wu2013, Moix2013, Fujita2014, Fujita2012, Liu2016, Dutta2016, Potocnik2018, Bondarenko2020, Li2022}.
The physical picture behind this treatment is that the electronically\hyphen coupled molecular units interact with a local environment causing a fast random modulation of their energy gaps. The open system dynamics of the density matrix is thus recovered as the average evolution over different realizations of the stochastic white noise or, equivalently, by solving a Markovian master equation resulting by assuming a memoryless environment with site dephasing couplings.

While the stochastic approach has intrinsic limits, as it leads to a high\hyphen temperature equilibration when used to modulate the site energies, it offers an ideal testbed to study non\hyphen Markovian dynamics in a conceptually clear setting. Indeed, the stochastic picture with classical noise has been widely used in chemistry and physics to capture lineshapes in optical and magnetic spectroscopy, as pioneered by Anderson and Kubo \cite{W.Anderson1954, Kubo1954}, decoherence in many\hyphen body and molecular systems \cite{Chenu2017, Gu2019}, and noise\hyphen assisted charge and energy transport \cite{Liu2016, Dutta2016, Pfluegl2000, Dijkstra2015, Fujita2012, Giusteri2017, Coccia2018}.
In this setting, an arbitrary correlation time of the environment can be introduced in the model simply by replacing white\hyphen noise fluctuations with a more general stochastic process.

Recently, the potential of quantum computers for simulating quantum dynamics has been intensively investigated, building on the early work by Lloyd demonstrating a substantial quantum advantage for dynamical simulations governed by any local Hamiltonian \cite{Lloyd1996a}.
For closed systems, Lloyd employed first-order Trotter decomposition to divide the unitary evolution map into smaller unitary operators propagating local contributions \cite{Lloyd1996a}. In addition to Trotterization, which is still the most popular approach for computing dynamics because of its versatility \cite{Jaderberg2022, Gallina2022, Bruschi2024, Poulin2011, Reiher2017, Sawaya2020}, alternative techniques employing, e.g., time-dependent variational principles \cite{Li2017, Yuan2019, Heya2023, Endo2020}, (truncated) Taylor expansion \cite{Berry2015} or the quantum power method \cite{Childs2021}, have been introduced during the last decades.
When moving from closed to open quantum systems, the unitarity of time evolution is lost, and translating the problem in terms of quantum (unitary) gates and qubit measurements becomes less trivial. A possible approach is to fully rely on the quantum advantage of simulating many-body dynamics and include explicitly a conveniently formulated environment into the simulation \cite{Guimaraes2024}. 
Yet, strategies to account implicitly for the role of the environment are highly valuable as they reduce the number of degrees of freedom to be treated explicitly in the Hamiltonian propagator. Indeed, several schemes have been proposed for simulating open system dynamics described by an arbitrary Markovian master equation \cite{Wang2011, Kamakari2022, Endo2020, Hu2020, Schlimgen2022} and specifically applied to the problem of excitonic transport \cite{Gupta2020, Mahdian2020a, Hu2021}.

In this context, we already discussed how digital quantum simulations of ENAQT dynamics can be performed by leveraging the unraveling of the Markovian master equation in terms of stochastic trajectories of pure states \cite{Gallina2022}. The key idea is to harness the repetitions of the quantum circuit, a necessary step in building measurement statistics in quantum computing, by incorporating suitable stochastic terms into the quantum circuits to simulate open system dynamics. In this paper, we generalize this idea to step out of the Markovian regime and investigate the role of memory effects on exciton transport by employing quantum trajectories which can be translated into quantum circuits.
Notice that random modulation of the control fields has been applied to analog quantum simulation of excitonic ENAQT in different quantum platforms such as nuclear magnetic resonance \cite{Wang2018}, trapped atomic ions \cite{Maier2019} and superconducting circuits \cite{Potocnik2018}.
The technique of including random modulations of the Hamiltonian into the quantum circuits is supported by the same considerations but with the advantage of the programmability of digital devices ensuring a straightforward encoding of different energetics and network topologies.   
Indeed, the design of quantum circuits based on trajectories modulated by colored noise leads us to analyze the resulting exciton dynamics and transport efficiency under different conditions.

The issue of memory effects on the efficiency of energy transfer has been debated especially in the context of photosynthetic molecular aggregates. Several studies supported a beneficial effect of non-Markovian dynamics for enhancing transport \cite{Fujita2012, Cui2021, Moreira2020, Rebentrost2009a}.
In this framework, it is tempting to associate an enhanced inter-molecule coherence time with a more efficient exciton transport \cite{Ishizaki2009a}.
However, other studies emphasized the robustness of the transfer efficiency in photosynthetic aggregates, meaning that memory effects were found not significant in determining the overall functionalities of the molecular aggregates \cite{Jesenko2013, Mujica-Martinez2013}. Our analysis highlights that the effects of correlation time are not ``universal" as its influence on efficiency depends on a delicate balance between Hamiltonian parameters and the incoherent contributions to the dynamics.
As a result, we identify a regime of ``memory-assisted” quantum transport which can be realized only in connection with a specific structure of the energy spectrum determined by the system Hamiltonian.

The remainder of the paper is organized as follows. Section \ref{sec:2} discusses the model used to study the dynamics when the local energy fluctuations are described by an Ornstein-Uhlenbeck process \cite{Gillespie1996}.
The equations of motion derived in Section \ref{subsec:to average dynamics} can be used both in a standard numerical implementation or for the synthesis of parametrized circuits for a quantum simulation (Section \ref{subsec:Parametrized quantum circuits}).
Although the quantum algorithm approach might allow scaling up the dimension of the problem in future quantum computers, it relies on Trotterization, which goes beyond the limits of current devices subject to (relatively) fast decoherence.
For this reason, in Section \ref{subsec:Quantum emulation}, we emulate the circuit execution classically to comment on the effect of Trotterization and finite measurement statistics that characterize the outcome of the quantum algorithm.
In Section \ref{sec:3}, we turn our attention to the excitonic dynamics emerging from the underlying stochastic Hamiltonian, and study transport efficiency as a function of the environment correlation time in a model system of four coupled chromophores in the whole range of system-environment coupling strengths.
The results of this Section are obtained by solving the dynamics with standard numerical techniques, not requiring Trotterization and sampling, which represent an overhead in a classical implementation.
In the concluding Section, we summarize the main findings and report final considerations.

\section{\label{sec:2}Digital quantum simulation of exciton transport with colored noise}

\subsection{Mapping the molecular network into a quantum register}

The energy transport problem can be studied in first approximation by specifying a network of two-level systems, representing the ground $\ket{g} \equiv \ket{0}$ and the excited $\ket{e} \equiv \ket{1}$ state of the $i$-th chromophore with excitation energy $\epsilon_{i}$.
Pairwise Coulombic interactions between chromophores are characterized by coupling parameters $V_{ij}$ that we assume to be real. In this setting, the Frenkel exciton Hamiltonian of a network of $N$ chromophores can be written as
\begin{equation} \label{eq:H_phys}
    H_{\text{phys}} = -\frac{1}{2} \sum_{i=1}^{N} \epsilon_{i} \sigma^{z}_{i} + \sum_{i=1}^{N-1} \sum_{j = i+1}^{N} V_{ij} \left(\sigma^{+}_{i} \sigma^{-}_{j} + \sigma^{-}_{i} \sigma^{+}_{j}\right),
\end{equation}
where $\sigma^{\pm} = \left( \sigma^{x} \mp i \sigma^{y} \right)/2$, while $\sigma^x$, $\sigma^y$ and $\sigma^z$ are the Pauli operators.

There are different ways of embedding such a system into a quantum register. For example, a straightforward mapping of the molecular network is obtained by associating each chromophore with a qubit of the quantum register. We refer to this embedding as the ``physical" encoding \cite{Gallina2022, Hines2007}, also known as one-hot or unary encoding \cite{Sawaya2020, DiMatteo2021}. The physical mapping requires $N$ qubits to simulate $N$ chromophores, with the dimension of the Hilbert space growing exponentially as $\dim(\mathcal{H}) = 2^{N}$. Such a representation includes all the manifolds of states characterized by a different number of excitations, from the ground state (no excitations) to the $N$-excitation state. This can be useful in addressing multiple excitations in the same network, as required, e.g., in the simulation of nonlinear electronic spectroscopies \cite{Bruschi2024, Bruschi2022, Bruschi2023, Bolzonello2023}.

However, this abundance of electronic states exceeds the active space involved in exciton transport in photosynthetic complexes under natural light conditions which typically involves only the single-exciton manifold. As a result, an exponentially large part of the Hilbert space remains unexplored during the dynamics generated by the Hamiltonian \ref{eq:H_phys} which conserves the number of excitations. To optimize the use of quantum resources (i.e., to reduce the number of qubits required), one can focus on the single-exciton Hamiltonian
\begin{equation}
    H_{\text{alg}} = \sum_{i=1}^{N} \epsilon_{i} \ketbra{i}{i} + \sum_{i=1}^{N-1} \sum_{j=i+1}^{N} V_{ij} \left( \ketbra{i}{j} + \ketbra{j}{i} \right),
\end{equation}
where $\ket{i}$ corresponds to the site-excitation state $\ket{i} \equiv \ket{g_{N} \dots e_{i} \dots g_{1}} \equiv \ket{g}_{N} \otimes \dots \otimes \ket{e}_{i} \otimes \dots \otimes \ket{g}_{1} $. Single-excitation states can be mapped to the (qubit) computational basis set through a resource-efficient binary encoding, requiring only $n = \lceil \log_{2} N \rceil$ qubits to implement a system with $N$ chromophores.
We refer to this mapping as the ``algorithmic'' encoding \cite{Gallina2022}, as the setting is commonly employed in quantum walk algorithms \cite{Hines2007, Shenvi2003}.

\subsection{From stochastic Hamiltonian\dots}

Within the stochastic Hamiltonian approach, we assume that the net effect of the coupling of the electronic states with other degrees of freedom is the random fluctuation of the site energies of the molecules in the network. From these fluctuations, the dynamics of the open system emerges as the average over noisy trajectories.

We focus on the site energy fluctuation $\delta \epsilon_{i}(t)$ while maintaining constant the inter-molecular coupling terms $V_{ij}$ \cite{Cui2021}.
The resulting time-dependent Hamiltonian of the system can be written as
\begin{equation}{\label{eq:stochastic_Hamiltonian}}
    H(t) = H + H^{\text{fluc}}(t),
\end{equation}
where
\begin{equation}
    H^{\text{fluc}}(t) = \sum_{i = 1}^{N} \delta \epsilon_{i}(t) F_{i},
\end{equation}
with $F_{i} = - \sigma^{z}_{i}/2$ in the case of physical mapping and $F_{i} = \ketbra{i}{i}$ for the algorithmic mapping. In the high\hyphen temperature limit, compared to the system bandwidth, energy fluctuations can be modeled as stochastic processes characterized by real-valued Time\hyphen Correlation Functions (TCFs). In particular, we assume site-energy fluctuations to be spatially\hyphen uncorrelated and to follow stationary Ornstein\hyphen Uhlenbeck (OU) processes characterized by the TCF
\begin{equation}{\label{eq:Ornstein-Uhlenbeck}}
    c_{\text{OU}}(t) = \left\langle 
\delta \epsilon_{i}(t) \delta \epsilon_{j}(0) \right\rangle = \delta_{ij} \frac{\Gamma}{\tau} e^{-\frac{\abs{t}}{\tau}},
\end{equation}
where $\Gamma/\tau$ is the amplitude and $\tau$ is the correlation time of the fluctuations, which quantifies the ``memory" of the environment. For the sake of simplicity, we assume that the amplitude and correlation time of the fluctuations are the same for each chromophore. The corresponding spectral function, defined as the Fourier Transform of the TCF, is the Lorentzian function
\begin{equation} \label{eq:lorentzian_SF}
    C_{\text{OU}}(\omega) = \int_{-\infty}^{+\infty} c_{\text{OU}}(t)e^{i \omega t} dt = 2 \Gamma \frac{1}{\omega^{2} \tau^{2} + 1}.
\end{equation}
This setting corresponds to a collection of independent overdamped Brownian oscillators coupled to the chromophores of the network \cite{Mukamel1995}, which is a commonly adopted model for the study of exciton transport \cite{Dijkstra2015,Bondarenko2020,Fujita2012,Fujita2014}. 

Notice that the OU process reduces to the well-known white-noise limit when the environment correlation time is shorter than the timescale of the system evolution.
The white noise is characterized by a $\delta$-correlation function and a flat frequency spectrum,
\begin{equation}
    c_{\text{W}}(t) = \lim_{\tau\rightarrow 0} c_{\text{OU}}(t) = 2 \Gamma \delta(t), \qquad C_{\text{W}}(\omega) = 2 \Gamma.
\end{equation}
In this regime, the resulting dynamics is well-described by the Haken-Strobl model \cite{Haken1973} and $\Gamma$ is the fundamental parameter defining the decoherence rate.
Quantum algorithms have already been proposed to tackle this situation \cite{Gallina2022,Hu2021}.

In the opposite limit of a slow environment, the ratio $\Gamma/\tau$ identifies the amplitude of the static disorder in the site energies, which is at the origin of Anderson-like localization phenomena \cite{Moro2018a, Moix2013, Somoza2017}.

\subsection{\label{subsec:to average dynamics}\dots  to average dynamics}
In quantum simulation, calculating the expectation value of an observable typically requires multiple measurements of the system state to obtain reliable statistics. For closed and isolated systems, it is sufficient to repeat the same quantum circuit several times. In our case, we will exploit the need for repetitions to implement different noise trajectories. Hence, we approximate the value of an open system observable as the average over the outcomes. To do so, the first step is to generate random trajectories of the energy fluctuations $\delta \epsilon_{i, \xi}(t)$ according to the TCF. From here on, the subscript $\xi$ indicates a specific noise realization. The initial condition is $\delta \epsilon_{i, \xi}(0) = n_{i, \xi}(0) \sqrt{\Gamma/\tau}$, where $n_{i, \xi}$ is a Gaussian random number with mean zero and variance one. The fluctuations at different times are obtained using the Gillespie updating formula \cite{Gillespie1996}
\begin{equation} \label{eq:fluctuation}
    \delta \epsilon_{i, \xi}(t + \Delta t) = \delta \epsilon_{i, \xi}(t) e^{- \Delta t/\tau} + n_{i, \xi}(t + \Delta t) \sqrt{\frac{\Gamma}{\tau} \left( 1 - e^{-2 \Delta t/\tau} \right)}.
\end{equation}
By inserting the obtained values in the fluctuating Hamiltonian (eq. \ref{eq:fluctuation}), we get a time-dependent Schr\"odinger equation describing the trajectory $\xi$ of the system state, whose integrated form reads
\begin{equation}{\label{eq:unitary_evolution}}
    \ket{\psi_{\xi}(t)} = U_{\xi}(t,0) \ket{\psi(0)} \qquad \text{with} \qquad U_{\xi}(t,0) = \underset{\leftarrow}{\mathcal{T}} e^{-i \int_{0}^{t} H_{\xi}(t) dt}
\end{equation}
where $\hbar=1$. Note that the solution can be straightforwardly generalized to an initially mixed state in the form $\rho(0) = \sum_{k} p_{k} \ketbra{\psi_{k}(0)}{\psi_{k}(0)}$, where $p_{k}$ is the probability of the system being in the state $\ket{\psi_{k}}$.

Due to the time-ordering, formal writing of the propagator $U_{\xi}(t)$ in eq. \ref{eq:unitary_evolution} does not have an analytical solution. For this reason, we discretize the time evolution into small time steps $\Delta t$
\begin{equation}
    U_{\xi}(t, 0) = \prod_{s=0}^{S-1} U_{\xi}((s+1) \Delta t, s \Delta t),
\end{equation}
where $S=T/\Delta t$ and we approximate
\begin{equation} \label{eq:Magnus_expansion}
    U_{\xi}((s+1) \Delta t, s \Delta t) \approx e^{-i \int_{s \Delta t}^{(s+1) \Delta t} H_{\xi}(t) dt},
\end{equation}
which corresponds to a first-order Magnus expansion \cite{Magnus1954, Iserles2000, Kormann2008,Tannor2006} where we are neglecting the effect of higher-order commutators. As the time step of the propagation becomes smaller, this truncated expansion becomes more accurate. A heuristic iteration of the dynamics with diminishing time step $\Delta t$ can be adopted to test the convergence. 

The solution of the integral over the fluctuations in eq. \ref{eq:Magnus_expansion} is analytical \cite{Gillespie1996}. However, since the time step is small, the integral can be approximated at first-order \cite{Gillespie1996}, giving 
\begin{equation} \label{eq:unitary_propagator}
    U_{\xi}((s+1) \Delta t, s \Delta t) \approx e^{-i H_{\xi}\left( s \Delta t \right) \Delta t}.
\end{equation}
This final form of the propagator corresponds \textit{de facto} to a constant energy value during the time step of evolution \cite{Bondarenko2020, Jansen2006}.
Remarkably, in the white-noise limit, this assumption cannot be valid, and the integral in eq. \ref{eq:Magnus_expansion} must be taken explicitly, resulting in the propagator discussed in ref. \cite{Gallina2022}.

From all the possible trajectories of the energy fluctuations, the exact expectation value of an observable $O(t)$ of the open system emerges as the average
\begin{equation} \label{eq:exact_expectation}
    \left\langle O(t) \right\rangle = \lim_{\Delta t \rightarrow 0} \overline{\left\langle O(t = S \Delta t) \right\rangle_{\xi}} = \lim_{\Delta t \rightarrow 0} \lim_{\Delta \Xi \rightarrow \infty} \frac{1}{\Xi} \sum_{\xi = 1}^{\Xi} \left\langle O(t = S \Delta t) \right\rangle_{\xi},
\end{equation}
where $\left\langle O(t) \right\rangle_{\xi} = \expval{O}{\psi_{\xi}(t)}$, while the overbar denotes the average over trajectories. Any finite $\Delta t$ (time discretization) and $\Xi$ (number of trajectories) introduce a numerical error. Therefore, the stochastic implementation only provides an estimation $\overline{\left\langle 
O(t=S\Delta t) \right\rangle_{\xi}}$ of the expectation value $\left\langle O(t) \right\rangle$.

\begin{figure*}
    \centering
    \includegraphics[width=\textwidth]{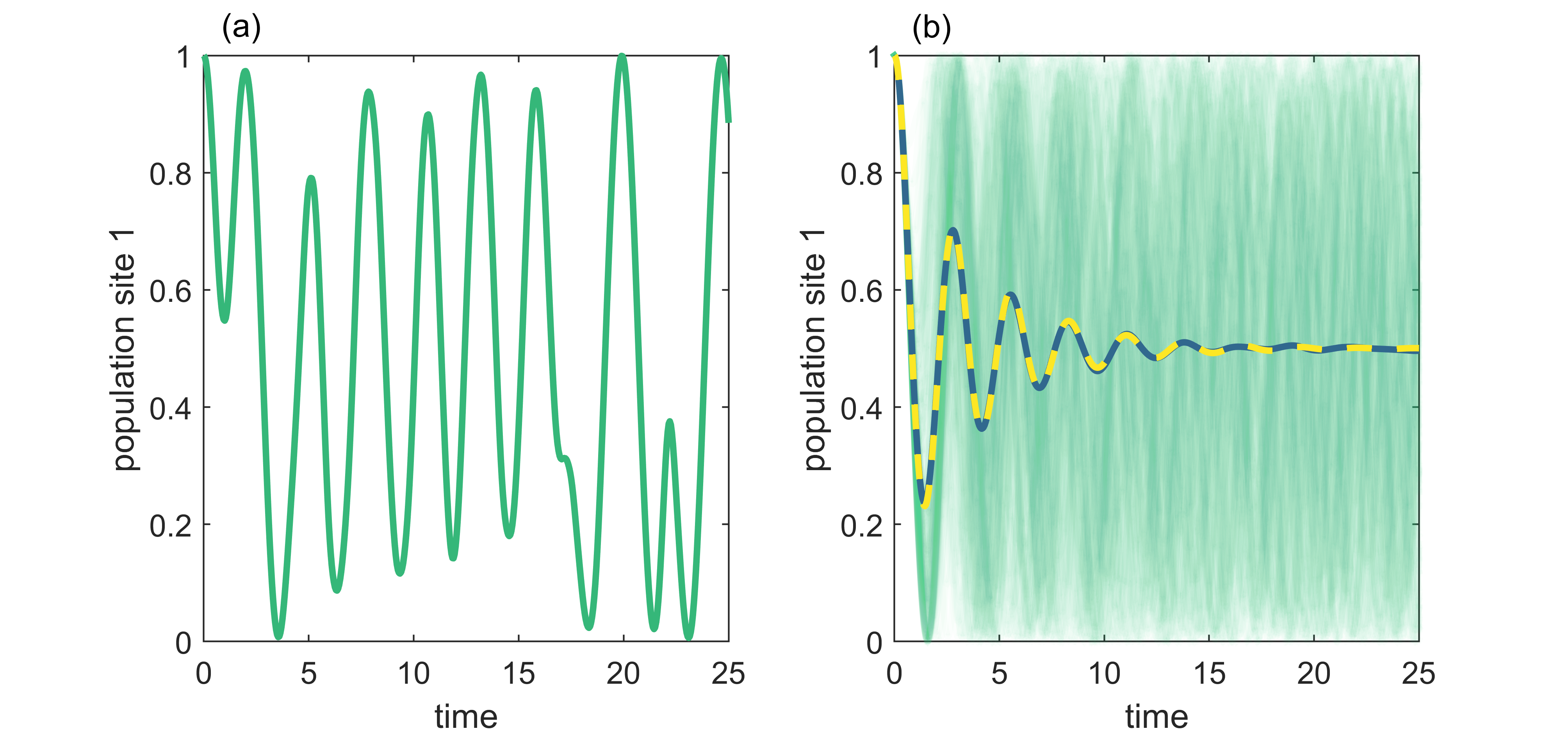}
    \caption{Dynamics of the population of the initially excited state of a homodimer system with site energy fluctuations ($\Gamma = 1$, $\tau = 1$). (a) Example of a single stochastic trajectory. (b) The open system dynamics (blue line) obtained as the average of $\Xi = 10^4$ trajectories (green lines in transparency) is compared with the HEOM benchmark (yellow dash line). Time evolution is carried out using the unitary propagator in eq. \ref{eq:unitary_propagator} with a time step $\Delta t = 0.05$.}
    \label{fig:trajectories}
\end{figure*}

In figure \ref{fig:trajectories}, we show how stochastic quantum trajectories appear in the case of energy transfer in an interacting homodimer.
From now on, we will set the inter-molecular coupling strength to $V = 1$. All energies, frequencies, and time units will be scaled accordingly.
Environment parameters are set to $\Gamma = 1$ and $\tau = 1$. Note that this situation corresponds to an intermediate system-environment coupling strength and a non-Markovian environment. Therefore, the evolution of the density matrix cannot be described by weak-coupling, Markovian master equations, such as Lindblad and Redfield equations.
We follow the excited-state population of site $i=1$, where the excitation was localized at the beginning of the dynamics.
In figure \ref{fig:trajectories}a, we display a single trajectory obtained with a random realization of the energy fluctuations. The evolution differs from the periodic oscillation of site populations of an isolated homodimer due to stochastic contributions. The overall behavior is similar to that reported in reference \cite{Gallina2022} for white noise, where the population changes smoothly over time.
Indeed, the stochastic Hamiltonian in eq. \ref{eq:stochastic_Hamiltonian} defines a particular instance of quantum state diffusion, where the quantum state is driven by a non-Markovian version of the stochastic Schrödinger equation \cite{Diosi1997,Diosi1998}. 
In figure \ref{fig:trajectories}b, we show the dynamics of $P_1(t)$ of the open system (solid blue line) obtained as the average over a swarm of $\Xi = 10^4$ trajectories (green lines in transparency) with different realizations of the energy fluctuations.
The consistency of the result is benchmarked by running a dynamics with the Hierarchical Equations Of Motion (HEOM, dashed yellow line) \cite{Lambert2023} using the same correlation function for the environment.
While individual trajectories display coherent beatings of site population because of their underlying unitary dynamics, the typical decoherence of the open system dynamics emerges from the loss of phase relation in the swarm of trajectories.

\subsection{\label{subsec:Parametrized quantum circuits}Parametrized quantum circuits}

Because of the unraveling of the open system dynamics through unitary stochastic trajectories, the quantum algorithm is readily implemented by the following steps: 
\begin{enumerate}[label=(\alph*)]
    \item Initialize the quantum register to encode the initial state of the system; 
    \item Generate the energy fluctuations $\delta \epsilon_{i}(s \Delta t)$ according to eq. \ref{eq:fluctuation};
    \item Implement the unitary propagator $U_{\xi}((s+1) \Delta t, s \Delta t)$ in terms of quantum gates acting on the quantum register;
    \item Repeat steps (b)-(c) for different time steps $s \in \{0, \dots, S-1 \}$;
    \item Measure the observable of interest at final time $\langle O(t = S \Delta t) \rangle_\xi$;
    \item Repeat steps (a) to (e) for the different noise trajectories $\xi \in \{ 1, 2, \dots, \Xi \}$;
    \item Take the average over the measurements to estimate $\overline{\left\langle O(S\Delta t) \right\rangle_\xi}$.
\end{enumerate}

To study quantum transport, the typical choice of the initial state consists of the excitation localized on a single chromophore $i$ of the network. Within the physical mapping, a localized state is created by applying an X-gate to the corresponding qubit of the quantum register. In the algorithmic mapping, the same state can be realized by applying an X-gate to those qubits being ``$1$" in the corresponding bitstring.

\begin{figure}
    \centering
    \includegraphics[width=\textwidth]{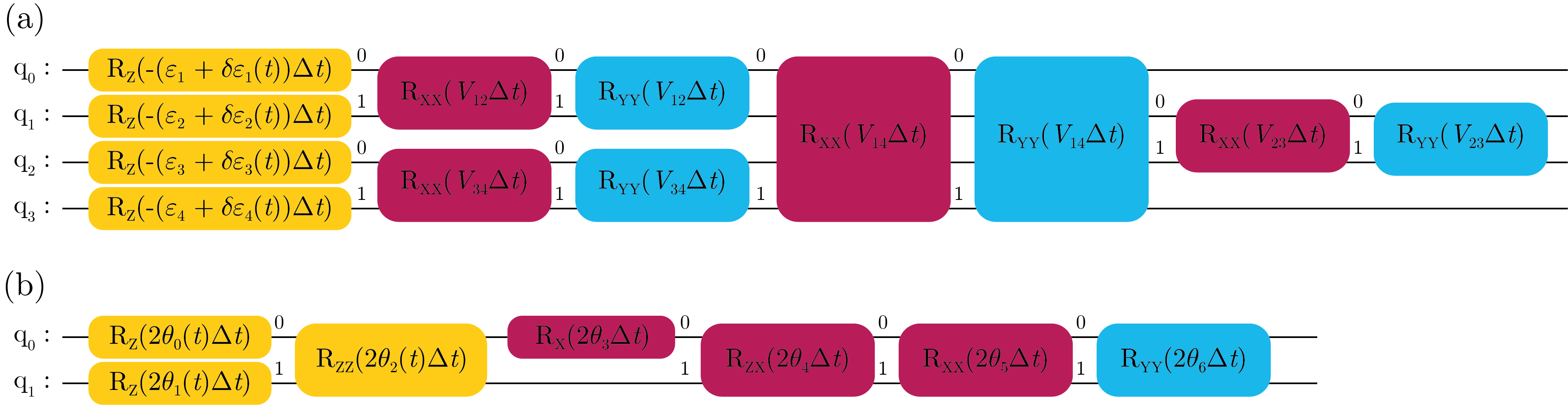}
    \caption{Quantum circuit for time propagator of a four-site cyclic network in physical (a) and algorithmic mapping (b). Energy fluctuations enter as (time-dependent) parameters for certain gates (see \ref{app:CNOT-staircase}). Circuits are reported for a Trotter number $m=1$. To augment the number of Trotter steps, $\Delta t$ should be replaced with $\Delta t / m$, and circuits should be repeated $m$ times.}
    \label{fig:circuits}
\end{figure}

The decomposition of the unitary dynamics in quantum gates, step (c), depends on the type of encoding used for the system. When the physical mapping is adopted, the propagator $U_{\xi}((s+1) \Delta t, s \Delta t)$ can be easily approximated using first-order Trotter decomposition as
\begin{equation} \label{eq:Trotter_physical}
    U_{\xi}((s+1) \Delta t, s \Delta t) \approx \left[ \left(\prod_{i=1}^{N-1}\prod_{j=i+1}^{N} e^{-i \frac{V_{ij}}{2} \sigma_{i}^{x} \sigma_{j}^{x} \frac{\Delta t}{m}} e^{-i \frac{V_{ij}}{2} \sigma_{i}^{y} \sigma_{j}^{y} \frac{\Delta t}{m}} \right) \left(\prod_{i=1}^{N} e^{i \frac{\epsilon_{i} + \delta\epsilon_{i}\left(s \Delta t\right)}{2} \sigma_{i}^{z} \frac{\Delta t}{m}} \right)\right]^{m},
\end{equation}
where $m$ is the Trotter number and we have used the equality $\sigma_{i}^{+} \sigma_{j}^{-} + \sigma_{i}^{-}\sigma_{j}^{+} = \frac{1}{2} \left( \sigma_{i}^{x} \sigma_{j}^{x} + \sigma_{i}^{y} \sigma_{j}^{y} \right)$ to rewrite the interaction terms of the Hamiltonian in eq. \ref{eq:H_phys}. The decomposed propagator can be straightforwardly translated into a quantum circuit made of one- and two-qubit gates $\text{R}_\text{Z}(\theta) = \exp(-i \theta \sigma^{z}/2)$, $\text{R}_\text{XX}(\theta) = \exp(-i \theta \sigma^{x} \otimes \sigma^{x}/2)$ and $\text{R}_\text{YY}(\theta) = \exp(-i \theta \sigma^{y} \otimes \sigma^{y}/2)$ as shown in figure \ref{fig:circuits}a for a four-site cyclic network.

If the problem is encoded with the algorithmic mapping, one can proceed by decomposing the Hamiltonian as a sum of Pauli strings \cite{Pravatto2021}, 
\begin{equation} \label{eq:Hamiltonian_to_Pauli}
\begin{split}
    H = \sum_{i,j=1}^{N} \mel{i}{H}{j} \bigotimes_{q=0}^{n-1} \bigg\{& \frac{1 - \abs{x_{q}(i) - x_{q}(j)}}{2} \left[ \mathbb{I}_{q} + (-1)^{x_{q}(i)} \sigma_{q}^{z} \right] +\\
    &+\frac{\abs{x_{q}(i) - x_{q}(j)}}{2} \left[ \sigma_{q}^{x} + (-1)^{x_{q}(i)} i \sigma^{y}_{q} \right] \bigg\},
\end{split}
\end{equation}
and then applying Trotterization together with the CNOT-staircase method to convert the propagator in a sequence of one- and two-qubit gates (see \ref{app:CNOT-staircase}). In figure \ref{fig:circuits}b, we report the quantum circuit for the time evolution for the same cyclic network in the case of algorithmic mapping. Gate parameters $\theta$s are given in \ref{app:CNOT-staircase}.

Notice that, in both physical and algorithmic mapping, the gate sequence is determined only by the connectivity of the molecular network dictated by the noiseless Hamiltonian in eq. \ref{eq:stochastic_Hamiltonian}.
Fluctuations influence only the angle of specific rotation gates, such as the $\text{R}_\text{Z}$ gates with the physical mapping, and one- and multi-qubit Z-rotation gates with the algorithmic mapping. Parametrized gates can therefore be exploited to speed up the quantum circuit composition.

In-depth scaling analysis of the quantum circuits is reported in \ref{app:scaling}, while the upper bounds are collected in table \ref{tab:scaling} for the worst-case scenario of a fully-connected molecular network. Despite requiring more qubits, the physical mapping exhibits some advantages in the number of operations performed during a Trotter step.

\begin{table}
    \centering
    \begin{tabular}{c|c|c|c}
        Mapping method & \# qubits & \# CNOTs (Trotter step) & circuit depth (Trotter step)\\
        \hline
        &&&\\
        Physical & $N$ & $\mathcal{O} \left( N^2 \right)$ & $\mathcal{O} \left( N \right)$ \\
        &&&\\
        \hline
        &&&\\
        Algorithmic & $\lceil \log_2 N \rceil$ & $\mathcal{O} \left( N^2 \log N \right)$ & $\mathcal{O} \left( N^2 \log N \right)$ \\
        &&&\\
        \hline
    \end{tabular}
    \caption{Upper bound scalings for the execution of a Trotter step with the number of chromophores $N$ in the aggregate. For the scaling, a molecular network with long-range interactions between all the chromophores is considered.}
    \label{tab:scaling}
\end{table}

\subsection{\label{subsec:Quantum emulation}Numerical simulation of the quantum circuits}
In this section, we use a classical computer as an emulator to run the quantum circuits and discuss the characteristics of the quantum algorithm. This corresponds to computing the Trotter dynamics for the system statevector and then performing a probabilistic sampling of the final state, simulating the quantum measurement.

\begin{figure*}
    \centering
    \includegraphics[width=\textwidth]{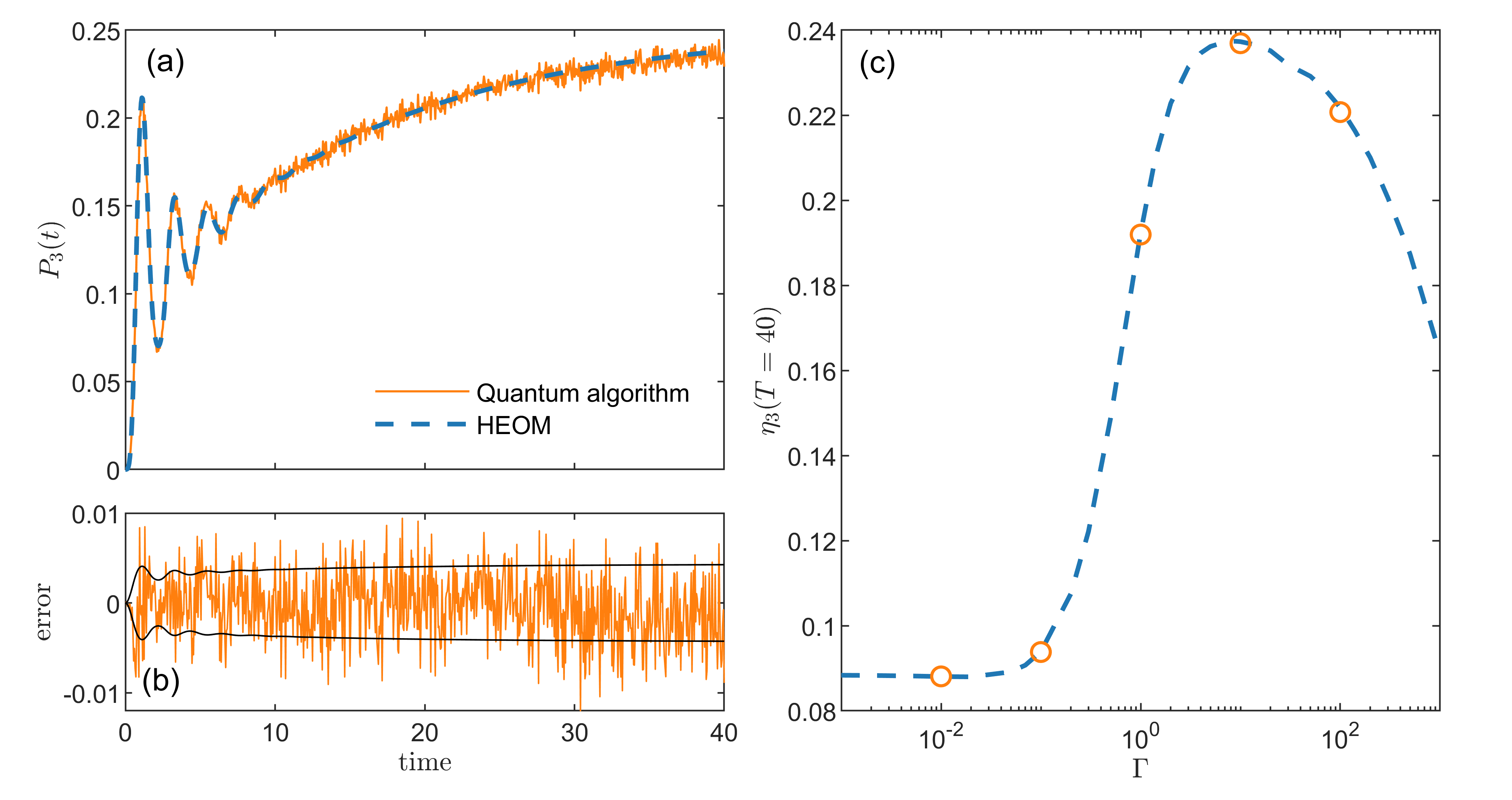}
    \caption{Classical emulation of the quantum circuits for a four-site cyclic network. (a) The evolution of the population at site 3 is obtained with the emulator (orange) and compared with HEOM (blue) for $\Gamma = 1$. (b) Their difference (orange) is reported along with the standard error of a binomial sampling with $K = 10^4$ trials (black). (c) By varying the value of $\Gamma$ at fixed $\tau$, the efficiency profile is calculated with the quantum emulator (orange scatters) and superposed to the classical average over $\Xi = 10^4$ trajectories (blue). In all panels, the memory time of the environment is $\tau = 1$, the time step for the algorithmic evolution is $\Delta t = 0.05$, and the number of Trotter steps per $\Delta t$ is $m = 3$.}
    \label{fig:quantum_emulation}
\end{figure*}

The results for a four-site cyclic network with a diagonal disorder are collected in figure \ref{fig:quantum_emulation}.
The site energies of the system have been sampled from a normal distribution with a variance of 4 in ref. \cite{Gallina2022}. Here we use the same Hamiltonian, given explicitly in eq. \ref{eq:Hamiltonian_example} (\ref{app:Hamiltonians}).
The system is supposed to have an initial excitation localized at site 1 and we observe population at site 3, $P_3(t)$, which is also used to calculate transport efficiency as discussed in Section \ref{sec:3}.
The memory time of the environment is $\tau = 1$.
The observable is obtained by averaging the outcome of $10^4$ independent circuits, each representing a unique realization of the noise trajectories.
We highlight that each circuit is sampled with a single measure (one shot) as discussed in \ref{app:measure}.

In figure \ref{fig:quantum_emulation}a, the dynamics of $P_3(t)$ from the quantum circuits (orange line) is compared with the HEOM solution (blue line) for $\Gamma = 1$.
As proof of convergence, we also report the deviation from the benchmark in figure \ref{fig:quantum_emulation}b (orange line), calculated as the difference between the populations at site 3 obtained with both methods.
Despite the effect of the finite sampling, the quantum algorithm faithfully reproduces the benchmark.
In fact, around 70\% of the time, the error falls within the range of $\pm \sqrt{P_3(t)\left( 1 - P_3(t) \right) / K}$ (black line),  which corresponds to the standard error of a binomial sampling with $K = 10^4$ trials.
This confirms that the error introduced by the numerical implementation is smaller than the sampling error on the benchmark solution.
Such a level of agreement is achieved by an accurate choice of the time step, which is set to $\Delta t = 0.05$, and the Trotter number per $\Delta t$, which is $m = 3$.
In general, the time step is bounded by the environment timescale and the fluctuation amplitude. The faster and larger the energy fluctuations, the smaller the time step.
On the other hand, the Trotter number is chosen to minimize the higher-order commutators during the Trotter decomposition, so it is required that the resulting $\Delta t / m$ is small compared to the evolution timescale of both the system and environment.
An appropriate choice of these two parameters is thus fundamental for the convergence of the results.

Furthermore, figure \ref{fig:quantum_emulation}c displays the value of the transport efficiency at different $\Gamma$ obtained by the quantum circuit emulator (orange dots) and by the average over $\Xi = 10^4$ trajectories generated classically without the use of Trotterization and finite sampling (blue line).
As already discussed, Trotterization can be governed by the choice of the time step, while the absence of visible finite sampling effects may be surprising.
However, when calculating time-averaged quantities, such as transport efficiency, the sampling noise that affects expectation values can be averaged away by taking enough measurement points along the dynamics. As a result, the quantum circuits estimate the target observable with an accuracy that is comparable to the classical implementation.
%

\section{\label{sec:3}Effects of the environment memory time on the system transport efficiency}
A question that naturally arises when considering colored noise is how different memory times influence the dynamical properties of an open quantum system. Here, we focus on transport efficiency and the emergence of ENAQT. We define transport efficiency as the mean population at a target site $j$ up to a threshold time $T$ \cite{Gallina2022,Maier2019,Li2022}
\begin{equation}\label{eq:efficiency}
    \eta_{j}(T) = \frac{1}{T} \int_0^T dt \, \rho_{jj}(t).
\end{equation}
For light-harvesting systems, this measure of transport efficiency is a good approximation when the exciton recombination and the leakage to other systems (e.g., a reaction center linked to the target site) are slow. All the numerical simulations reported in this section are performed with a classical implementation of the equations in Section \ref{subsec:to average dynamics}, which does not require gate decomposition and measurement sampling.

\subsection{\label{subsec:Effects of memory time on transport efficiency}Transport efficiency in a colored environment: typical scenario}

Our initial case study is the same four-site cyclic network with diagonal disorder introduced in Section \ref{subsec:Quantum emulation}, whose Hamiltonian is reported in eq. \ref{eq:Hamiltonian_example}.
To study the effect of memory, we simulate the system dynamics under different noise conditions.

\begin{figure*}
    \centering
    \includegraphics[width=\textwidth]{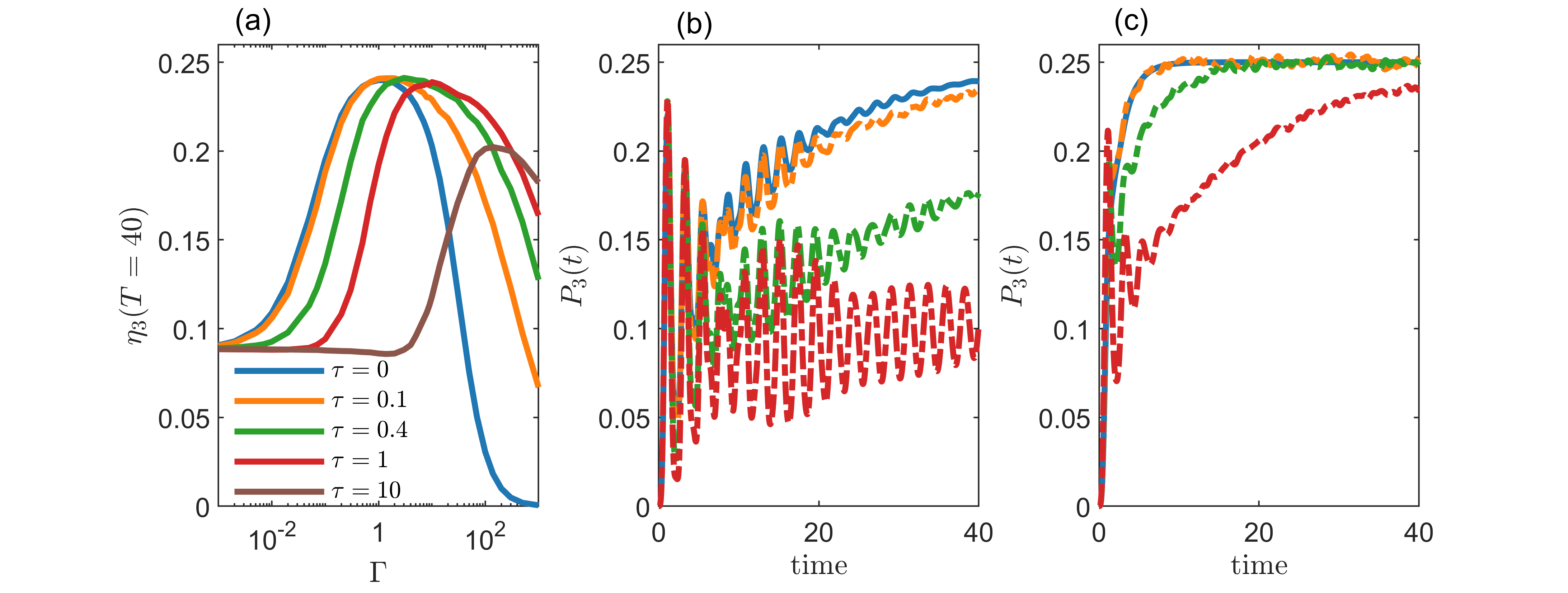}
    \caption{Transport efficiency in a four-site cyclic network. (a) Efficiency is shown for target time $T = 40$ as a function of the noise strength $\Gamma$. Dynamics of the target site population at fixed noise strength $\Gamma = 0.1$ and $\Gamma = 1$ are reported in panels (b) and (c) respectively. Differences in the environment correlation time, $\tau = 0$ (blue), $\tau = 0.1$ (orange), $\tau = 0.4$ (green), $\tau = 1$ (red) and $\tau = 10$ (brown), lead to different damping of the coherent beatings and relaxation timescale.}
    \label{fig:enaqt_gamma}
\end{figure*}

Figure \ref{fig:enaqt_gamma}a shows the transfer efficiency as a function of $\Gamma$ for different environment correlation times.
The noise parameter $\Gamma$ has a well-defined counterpart in a microscopic model of a bath made of overdamped Brownian oscillators at high temperature, $\beta \to 0$.
It reflects the ratio between the reorganization energy $\lambda$ and the cutoff frequency $\omega_\text{c}$, namely $\Gamma = 2 \lambda / \beta \omega_\text{c}$ \cite{Gallina2023, Bondarenko2020}.
In reference \cite{Mohseni2014}, this ratio is recognized as the parameter controlling ENAQT in the Fenna-Matthews-Olson (FMO) complex. In that case, the efficiency enhancement is found to be important when such a parameter is comparable to the average energy gap of the excitonic manifold.
By varying $\Gamma$, we can compare the dynamics including memory effects with the dynamics in the corresponding memoryless limit ($\tau = 0$, $\Gamma = \text{const.}$, see eq. \ref{eq:lorentzian_SF}).
Indeed, in the limit of vanishing memory time, $\Gamma$ quantifies the variance of the white-noise fluctuations and it defines the site decoherence rate according to the Lindblad equation corresponding to the Haken-Strobl model \cite{Gallina2022}
\begin{equation} \label{eq:Lindblad}
    \frac{d\rho(t)}{dt} =
    -i \left[ H, \rho(t) \right] +
    \sum_{i=1}^N 2 \Gamma \left( \ketbra{i} \rho(t) \ketbra{i} - \frac{1}{2} \left[ \ketbra{i}, \rho(t) \right]_+ \right),
\end{equation}
where $\left[ A, B \right]$ and $\left[ A, B \right]_+$ are the commutator and anti-commutator of operators $A$ and $B$, respectively.

In all cases, we observe the characteristic peak of efficiency, which is the hallmark of ENAQT. This trend has been observed for a large variety of networks and can be understood by considering the interplay between coherent and incoherent evolution. The understanding of ENAQT has been mainly developed by considering the dynamics in the Markovian (memoryless) regime, but the same considerations apply to the non-Markovian cases as shown in figure \ref{fig:enaqt_gamma}a.
In the region before the maximum efficiency, the environment weakly contributes to the system dynamics, and the transport efficiency is comparable with that of the isolated system, which is determined by the static disorder in the site energies. In the specific case of the Hamiltonian in eq. \ref{eq:Hamiltonian_example}, a prominent energy gap between site $3$ and its neighbors results in a limited overlap between the target site and the Hamiltonian eigenvalues which are highly populated at the beginning of the dynamics.
As a consequence, the Schr\"odinger dynamics of the isolated system is expected to transfer a limited amount of exciton population to the target site, and the system-environment coupling is not sufficient to overcome the partial localization within the threshold time $T$.
An example is the dynamics for $\tau = 1$ in figure \ref{fig:enaqt_gamma}b (red line), which resembles the evolution of the isolated system while it is slightly perturbed by the fluctuating environment.

By increasing $\Gamma$, transport efficiency is enhanced up to a central plateau, which extends for some orders of magnitude of $\Gamma$, representing the maximum efficiency.
Indeed, site-dephasing ultimately leads the system to an equilibrium state where all sites have equal populations $1/N$ (figure \ref{fig:enaqt_gamma}c). This population equipartition implies a higher probability of finding the exciton on site 3 compared to the isolated case.
When the equipartition timescale is comparable to the threshold time $T$, we observe an enhanced efficiency.

Finally, in the limit of high $\Gamma$, when the incoherent dynamics is the main actor, the excitonic coupling terms in the Hamiltonian can be treated as a perturbation and transport is suppressed.
In the Markovian case, the effect is the same as stroboscopic measurements of exciton localization, also called the quantum Zeno effect \cite{Rebentrost2009}.

While the overall behavior of the efficiency as a function of the noise strength $\Gamma$ is conserved moving outside the Haken-Strobl limit, finite-memory effects are evident in the drift of the ENAQT profiles going from the blue line (memoryless environment) to the brown line (long-time-correlated environment) in figure \ref{fig:enaqt_gamma}a.
In the region of weak-to-intermediate noise strength, $\Gamma/V \le 1$, a higher noise strength is required to induce enhancement of the efficiency when increasing the correlation time of the environment.
To rationalize this effect, let us analyze the population dynamics at the target site, site 3, for two different values of the noise strength $\Gamma = 0.1$ (figure \ref{fig:enaqt_gamma}b) and $\Gamma = 1$ (figure \ref{fig:enaqt_gamma}c) for different correlation time $\tau$.
In both cases, the Markovian dynamics is closer to the maximum efficiency, while the efficiency decreases as the environment gets slower.
Interestingly, for the same noise strength, longer memory times of the environment have the effect of sustaining the system coherences, as reflected by the persistent coherent beating of the site population (red and green traces).
Such coherent beatings will necessarily die out following the complete equilibration of the dynamics toward an equally distributed site population.
However, the timescale to reach the equipartition of population depends strongly on the environment correlation time $\tau$.
It is such an equipartition timescale, rather than long-living coherences, which determines the enhancement of efficiency in the region $\Gamma/V \le 1$ as we will further discuss below.
Therefore, higher values of the noise strength $\Gamma$ are necessary to induce a similar enhancement of the efficiency as highlighted by the right shift of the efficiency profile.

Let us now analyze the region of high noise strength $\Gamma/V > 1$, where memory enhances substantially the efficiency compared to the Markovian case, even when the correlation time is small (orange line in figure \ref{fig:enaqt_gamma}a). In this parametric region, the dynamics is largely controlled by the dissipative contributions and the excitation transport becomes substantially incoherent as described by F\"orster theory of energy transfer.
By treating the coupling $V_{ij}$ between sites as a perturbation, the evolution of the site populations is well-approximated by a hopping mechanism where the transfer rates are calculated via the Fermi golden rule \cite{Dijkstra2015,Wu2012,Wu2013,Ishizaki2009,Yang2002}.
In this case, the site population dynamics follows the kinetic equation
\begin{equation}
    \frac{dP_i(t)}{dt} =
    \sum_{j=1}^N \kappa_{ij} P_j(t), 
\end{equation}
where $P_i(t) = \rho_{ii}(t)$ and the transfer rates are determined by the Hamiltonian and the noise as \cite{Dijkstra2015,Wu2013}
\begin{equation} \label{eq:rates}
    \kappa_{i \neq j} =
    2 \abs{V_{ij}}^2
    \text{Re} \left\{
    \int_0^\infty
    dt \,
    e^{
    i \left( \epsilon_i - \epsilon_j \right) t -
    2 \Gamma t +
    2 \Gamma \tau \left( 1 - e^{-t/\tau} \right)
    }
    \right\}
\end{equation}
with $\text{Re}\{A\}$ the real part of $A$, and
\begin{equation}
    \kappa_{ii} = - \sum_{j \neq i} \kappa_{ij}.
\end{equation}
\begin{figure*}
    \centering
    \includegraphics[width=\textwidth]{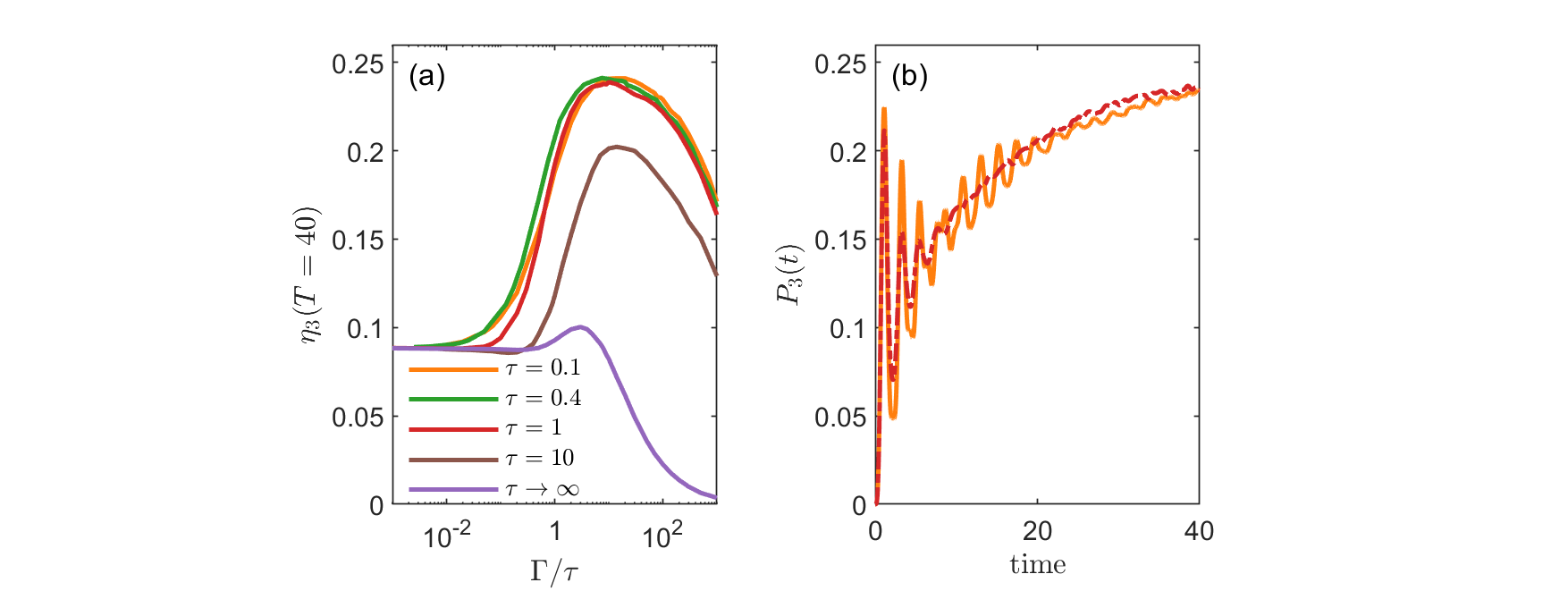}
    \caption{Transport efficiency in a four-site cyclic network. (a) Efficiency is shown as a function of the noise fluctuation amplitude $\Gamma/\tau$. Dynamics of the target site population (b) are reported at fixed fluctuation amplitude $\Gamma/\tau = 1$. Differences in the environment correlation time, $\tau = 0$ (blue), $\tau = 0.1$ (orange), $\tau = 0.4$ (green), $\tau = 1$ (red), $\tau = 10$ (brown) and $\tau \to \infty$ (violet), lead to similar rates toward equipartition of site population but different coherence-dephasing rates are experienced by the open system. For the static energy disorder (violet line) the dynamics is calculated as the average over $10^4$ Schr\"odinger propagations with time-independent stochastic energies.}
    \label{fig:enaqt_gamma_tau}
\end{figure*}
The site-to-site transfer rates decrease when $\Gamma$ becomes large, causing the efficiency to decrease.
However, for a given $\Gamma$, a longer memory time leads to higher transfer rates and thus higher efficiency in agreement with the results shown in figure \ref{fig:enaqt_gamma}a.
In particular, in the Markovian limit, the transfer between two sites with the same energy is proportional to $\Gamma^{-1}$ while, in the limit $\tau \gg t$, the same rate is proportional to the inverse of the fluctuation amplitude, $\left( \Gamma/\tau \right)^{-1/2}$.
By examining the efficiency profiles as a function of the fluctuation amplitude, $\Gamma/\tau$, we notice that the efficiency profiles for different memory times are well-superposed (see figure \ref{fig:enaqt_gamma_tau}a).
This suggests that, for a wide range of $\tau$, the fluctuation amplitude $\Gamma/\tau$ is the only relevant parameter to determine the efficiency.
By looking at the dynamics with different memory times but with the same noise amplitude $\Gamma/\tau$ (figures \ref{fig:enaqt_gamma_tau}b), we can see that the memory time $\tau$ relates to the persistence of the coherent beatings of the site population.
However, the rate toward population equipartition remains similar because this is controlled by the fluctuation amplitude as long as the correlation time remains comparable to the timescale of the Hamiltonian evolution.
For a given noise amplitude, increasing the memory of the environment reduces the coherent beatings.

Another regime is encountered for correlation time longer than the characteristic time of the system evolution. In this case, the efficiency enhancement is limited due to static disorder and populations display a pre-equilibrium before reaching equipartition at a longer timescale. In the limit $\tau \to \infty$, the dynamics is calculated as the average over $10^4$ unitary propagations with time-independent stochastic energies, i.e. static disorder (violet line). Interestingly, the efficiency profile still reports a maximum value at an intermediate variance of the static disorder suggesting that the Hamiltonians built with typical random energies are characterized by a higher mixing of the target site with the initial site than the specific reference Hamiltonian, eq. \ref{eq:Hamiltonian_example}.

\subsection{\label{subseq:Memory-assisted quantum transport}Memory-assisted quantum transport}
\begin{figure*}
    \centering
    \includegraphics[width=\textwidth]{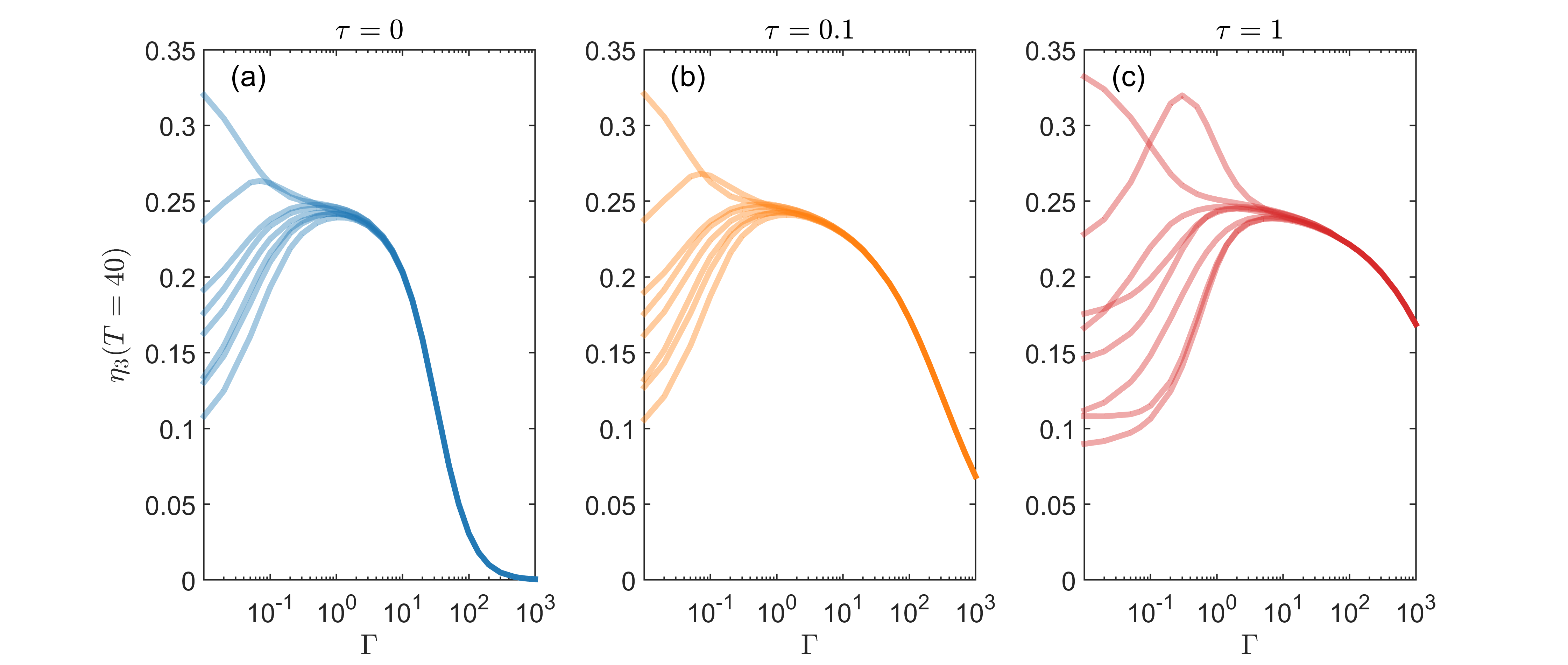}
    \caption{Transport efficiency in different realizations of the site-energy static disorder in four-site cyclic networks. Different memory times of the environment are reported: $\tau = 0$ (a), $\tau = 0.1$ (b) and $\tau = 1$ (c).}
    \label{fig:enaqt_vari}
\end{figure*}

In the four-site example discussed above, finite memory does not allow for higher efficiencies than those achievable under Markovian conditions.
Because efficiency stands from a delicate balance between coherent and incoherent evolution, an interesting question is whether other Hamiltonians admit a different phenomenology where finite correlation time in the environment improves transport efficiency.
To answer this question, we generated a pool of random Hamiltonians with energy disorder drawn from the aforementioned Gaussian distribution with a variance of 4. We analyze the profiles of transport efficiency in the memoryless limit (figure \ref{fig:enaqt_vari}a) and for increasing values of the environment correlation time (figure \ref{fig:enaqt_vari}b,c).

\begin{figure*}
    \centering
    \includegraphics[width=\textwidth]{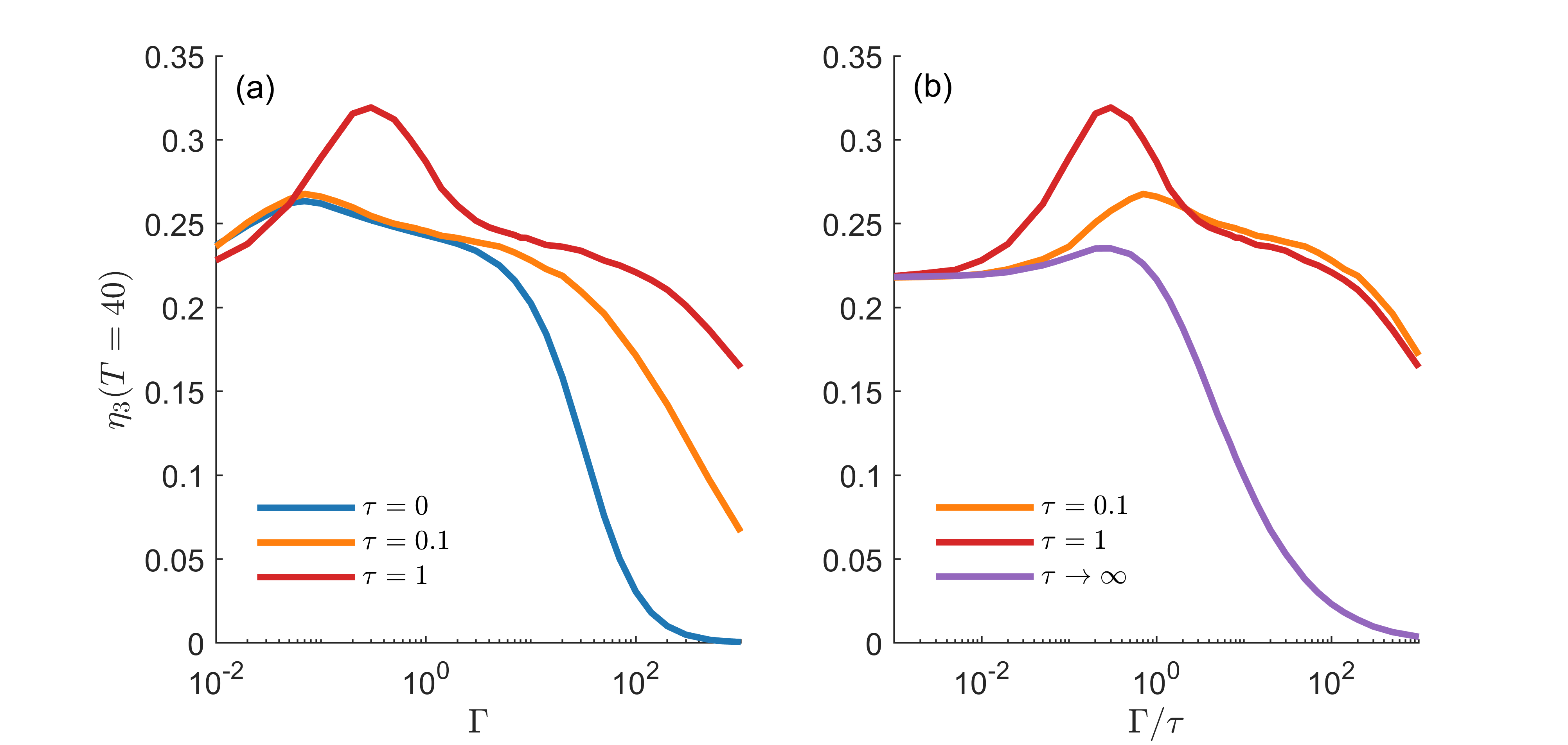}
    \caption{Transport efficiency in a four-site cyclic network showing a peak enhanced by an intermediate memory time for the environment. Efficiency profiles are shown against the noise strength $\Gamma$ (a) and noise fluctuation amplitude $\Gamma/\tau$ (b).}
    \label{fig:enaqt_MAQT}
\end{figure*}

As can be noted, the phenomenology of a single broad maximum in efficiency at values of noise strength comparable to the Hamiltonian coupling is rather typical.
However, two realizations present different behaviors.
In one case, the efficiency decreases monotonically with the noise strength and does not shows any environment-induced enhancement.
This happens when the network is (almost) ordered and therefore the coherent ballistic transport assures the highest efficiency \cite{Kassal2012}.
In another case, the system exhibits improved maximum efficiency in the presence of memory effects.
The Hamiltonian for this specific system is reported in eq. \ref{eq:Hamiltonian_MAQT} (\ref{app:Hamiltonians}), while details of the efficiency profiles are reported in figure \ref{fig:enaqt_MAQT} as a function of the noise strength (figure \ref{fig:enaqt_MAQT}a) and fluctuation amplitude (figure \ref{fig:enaqt_MAQT}b). Notice that the memory-enhanced peak in the efficiency profile (red line) is located before the shoulder where the system passes from the coherent to the incoherent regime, typically corresponding to the maximum-efficiency plateau in disordered systems.

By looking at the Hamiltonian eigenvector matrix in eq. \ref{eq:eigenvectors_MAQT}, an efficient mixing is expected between the initial state, localized at site 1, and the target site 3. As a consequence of this mixing of the two sites, the population of the target site quickly grows above the equipartition value ($P = 1/4$) during the coherent dynamics.

\begin{figure*}
    \centering
    \includegraphics[width=\textwidth]{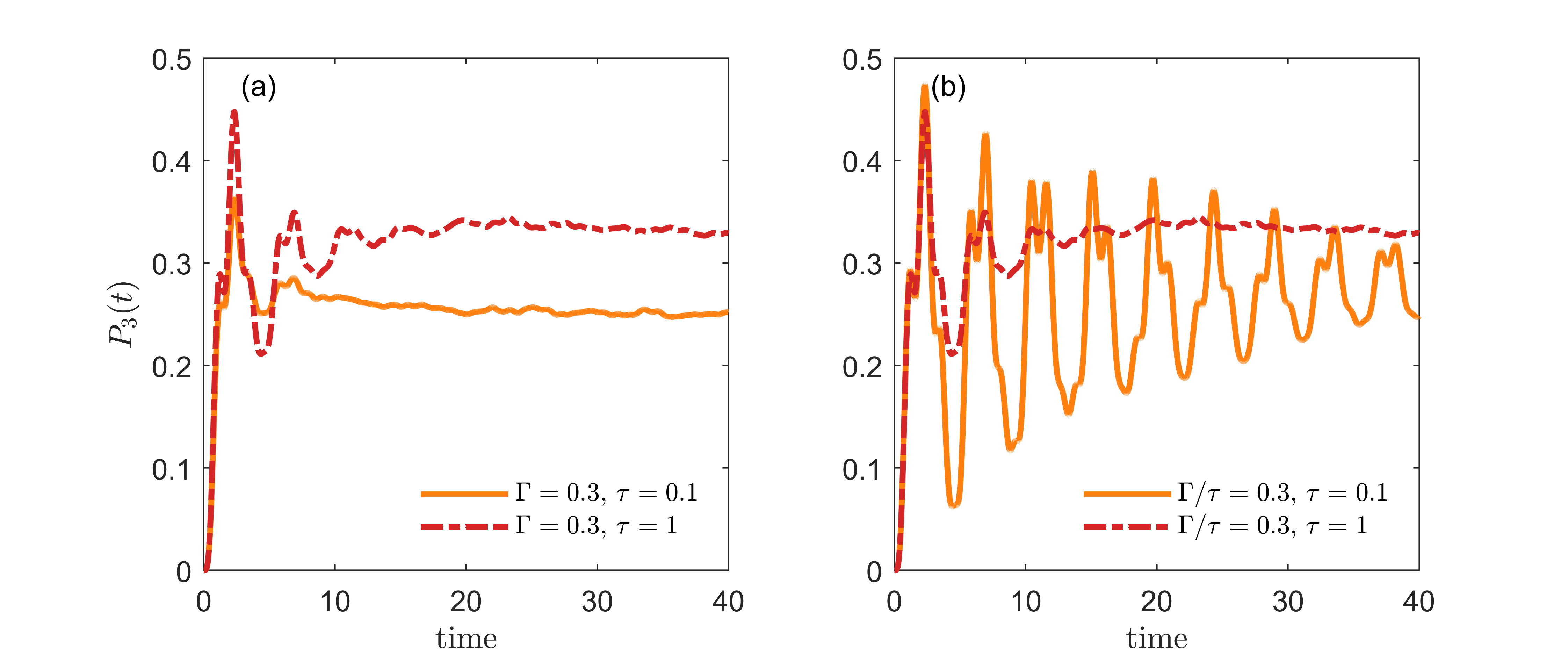}
    \caption{Comparison of the dynamics of the target site population for environments with noise strength $\Gamma = 0.3$ (a) and noise fluctuation amplitude $\Gamma/\tau = 0.3$ (b). Differences in the memory time of the environment, $\tau = 0.1$ (orange) and $\tau = 1$ (red), lead to different relaxation timescales in (a) and different damping of the coherent beatings in (b).}
    \label{fig:MAQT_dinamiche}
\end{figure*}

In figure \ref{fig:MAQT_dinamiche}, we show the dynamics of the population of the target site for the value of $\Gamma$ (figure \ref{fig:MAQT_dinamiche}a) and $\Gamma/\tau$ (figure \ref{fig:MAQT_dinamiche}b) corresponding to the maximum transport efficiency for $\tau = 1$.
The two lines refer to different memory times, $\tau = 0.1$ and $\tau = 1$.
At fixed $\Gamma$, we observe the same general phenomenology as in the previous section, namely longer memory times result in a slower damping of the coherent beating and a slower energy-equipartition rate.
However, with the new set of site energies in eq. \ref{eq:Hamiltonian_MAQT}, this results in increased efficiency in a strongly non-Markovian environment.
This is due to the interplay with the Schr\"odinger dynamics of the isolated system presenting coherent oscillation of the target population with maximum value above the equipartition.
In this particular case, a slower damping, which causes a longer permanence of population at site 3 results in an increased efficiency.
By comparing the dynamics at fixed $\Gamma/\tau$ (figure \ref{fig:MAQT_dinamiche}b), we also observe that for a similar equipartition rate, the faster damping of the coherent beatings, due to longer correlation time, favors efficiency by stabilizing the population on the target site 3.

As a final note, we briefly comment on the efficiency profiles of figures \ref{fig:enaqt_vari} and \ref{fig:enaqt_MAQT} in the region of high noise strength.
In this regime, eq. \ref{eq:rates} suggests that above a threshold value of $\Gamma$, the contribution of site energies to the kinetic rates becomes irrelevant.
Above such a critical value, only the network topology, $V_{ij}$, and the noise parameters, $\Gamma$ and $\tau$, determine the transfer rates while the Hamiltonian part becomes irrelevant as demonstrated in figure \ref{fig:enaqt_vari}. 
%
\section{\label{sec:Conclusions}Conclusions}
In this work, we have elaborated on the dynamics of exciton transport in a molecular network in non-Markovian conditions within the well-defined framework of stochastic Hamiltonians.
While the generalization of master equations beyond the Markovian and weak-coupling assumptions leads to significant complexity of the description, the study of the dynamics generated by a fluctuating environment can be handled as the Schr\"odinger equation for the isolated system and it offers a straightforward unraveling of the open system dynamics.
By assuming site energy fluctuations according to an Ornstein–Uhlenbeck process, we first discuss quantum circuits that are suited to simulate the open system dynamics for a wide range of coupling strengths and correlation times.
The simulation strategy, based on the unraveling of the open system dynamics, generalizes our previous work on the digital quantum simulation of environment-assisted quantum transport in the Markovian regime \cite{Gallina2022} and it enriches the library of quantum algorithms available for the simulation of non-Markovian dynamics \cite{Wang2023,Head-Marsden2021}.
In particular, while algorithms based on dilation of the Hilbert space require a priori knowledge of the open system dynamics to be simulated, the underlying idea of the presented approach is to encode an effective representation of the environment and thus obtain the open system dynamics as resulting from the average over unitary trajectories.
We demonstrated that the quantum circuits for the Trotterized Hamiltonian propagators are common to networks with the same connectivity, and only specific rotation gates depend on the stochastic realizations of energy fluctuation. This feature offers the opportunity to use parametrized quantum circuits to speed up the circuit compilation. Moreover, a quantum device that optimizes the execution of circuits made of the same quantum gates with varying parameters could further speed up the computation in a real execution of the quantum algorithm.
Another direction for optimizing the algorithm efficiency is the study of higher-order Magnus expansion in eq. \ref{eq:Magnus_expansion}, possibly allowing the choice of a larger $\Delta t$ \cite{Ture2024}.
In all cases, the Trotterization of the evolution operator generates rather long circuits requiring to preserve high fidelity over the execution time.
Recently, the idea of harnessing the hardware noise in quantum circuits has been explored \cite{Tolunay2023, Sun2024, Guimaraes2023} and its application in the context of stochastic propagators is an intriguing possibility.
However, manipulating the intrinsic noise of digital quantum computers is a hard task and its conversion into a programmable stochastic process has not yet been achieved.

On the other hand, we also studied in detail the effect of memory times on transport efficiency emerging from the underlying model. Here, the dynamics of a disordered cyclic network with four sites has been taken as a reference. We showed that typically, i.e., in most of the cases when different realizations of the static site energies are taken into account, the efficiency profile results to be robust over a wide range of memory times and coupling with the environment, which is compatible with observations of the robustness of environment-assisted quantum transport \cite{Mohseni2014}.
However, we noted that, for certain systems, an increase in correlation times of the environment leads to the enhancement of transport efficiency.
We trace back this evidence to a strong interplay between the coherent and the incoherent contributions to the quantum dynamics in determining the overall efficiency. While design guidelines for efficient transport have been recognized for unitary and Markovian evolution (see, e.g., \cite{Mattiotti2022, Zech2014}), engineering of the Hamiltonian that allows for the maximization of memory-enhanced transport is still missing and it is an intriguing issue for future works.
A rationalization of this effect could be useful for the design of optimized networks for energy transport.

Notice that the dynamics of the four-site model system has been simulated both with a quantum emulator running the quantum circuits (Section \ref{subsec:Quantum emulation}) and with an analogous classical implementation which does not require the gate decomposition of the stochastic propagators nor measurement sampling. A rather general and relevant point to discuss is when a quantum computing approach becomes more convenient than an analogous classical implementation. A first aspect relates to the dimension of the chromophore network because of the quantum memory advantage. Indeed, while the classical storage of a quantum state grows linearly with $N$, the number of qubits in the algorithmic setting scales as $\log_2(N)$ (see table \ref{tab:scaling}). On the other hand, the algorithmic complexity for a simple excitonic network is comparable in the quantum and the classical case considering that a classical matrix-vector product requires $O(N^2)$ elementary operations, like the number of CNOTs for a single Trotter step. However, a substantial advantage of a quantum implementation in terms of complexity is expected for simulations of quantum dynamics ruled by a many-body Hamiltonian. In exciton systems, this is the case when considering multi-exciton dynamics or upon including exciton-vibration coupling into the Hamiltonian (see, e.g., ref. \cite{Bruschi2024}). Simulations of these more complex systems will be the real added value of a mature quantum computing technology.
\ack
The authors acknowledge the financial support by the Department of Chemical Sciences (DiSC) and the University of Padova with Project QA-CHEM (P-DiSC No. 04BIRD2021-UNIPD). B.F. acknowledges funding from the European Union - NextGenerationEU, within the National Center for HPC, Big Data and Quantum Computing (Project no. CN00000013, CN1 Spoke 10: ``Quantum Computing'').
The authors acknowledge the C3P facility of the Department of Chemistry of the University of Padova for the computing resources for the simulations.
We acknowledge the CINECA award under the ISCRA initiative, for the availability of high-performance computing resources and support.

\appendix
\section{CNOT staircase} \label{app:CNOT-staircase}
Adopting an algorithmic mapping poses the problem of decomposing Hamiltonian evolution in terms of gates.
We start by writing the Hamiltonian as a sum of products of Pauli operators (so-called Pauli strings).
This can be done by considering that any operator can be written as $H = \sum_{ij} H_{ij} \ketbra{i}{j}$, where $H_{ij} = \bra{i} H \ket{j}$.
Vectors $\ket{i}$ can be expanded over the computational basis set of the quantum register, $\ket{i} \equiv \ket{x_{n-1}(i)} \otimes \dots \otimes \ket{x_0(i)}$, so that 
\begin{equation}
    H = \sum_{ij} H_{ij} \bigotimes_{q=0}^{n-1} \ket{x_q(i)}\bra{x_q(j)},
\end{equation}
where the tensor product runs on the qubits (indexed by the letter $q$) and $x_q(i)$ can be either 0 or 1.
Each single-qubit term can be expressed as a combination of Pauli operators, following the scheme
\begin{equation}
    \begin{split}
        &\ketbra{0} = \frac{1}{2} \left( \mathbb{I} + \sigma^z \right) \\
        &\ketbra{0}{1} = \frac{1}{2} \left( \sigma^x + i\sigma^y \right) \\
        &\ketbra{1}{0} = \frac{1}{2} \left( \sigma^x - i\sigma^y \right) \\
        &\ketbra{1} = \frac{1}{2} \left( \mathbb{I} - \sigma^z \right). \\
    \end{split}
\end{equation}
The resulting Hamiltonian has the form
\begin{equation} \label{app_eq:pauli_string_H}
    H = \sum_{k} h_k \Pi_k
\end{equation}
where $\Pi_k = \bigotimes_{q=0}^{n-1} \sigma_q (k)$ are called Pauli strings.

For example, by using a standard binary mapping where state $\ket{i}$ of the system is encoded into state $\ket{\text{bin}(i-1)} = \ket{x_{n-1}(i) \dots x_{0}(i)}$ of the qubit register, the Hamiltonian of the four-site cyclic network (figure \ref{fig:circuits}) becomes: for the diagonal part
\begin{equation}
\begin{split}
    H^\text{d} =& \epsilon_1(t) \ketbra{1} + \epsilon_2(t) \ketbra{2} + \epsilon_3(t) \ketbra{3} + \epsilon_4(t) \ketbra{4}\\
    \xrightarrow{\text{to qubits}}&
    \epsilon_1(t) \ketbra{00} + \epsilon_2(t) \ketbra{01} + \epsilon_3(t) \ketbra{10} + \epsilon_4(t) \ketbra{11}\\
    \xrightarrow{\text{to Pauli}}&
    \theta_{-1}(t) \mathbb{I} \otimes \mathbb{I} +
    \theta_{0}(t) \mathbb{I} \otimes \sigma^z +
    \theta_{1}(t) \sigma^z \otimes \mathbb{I} +
    \theta_{2}(t) \sigma^z \otimes \sigma^z
\end{split}
\end{equation}
where $\epsilon_i(t) = \epsilon_i + \delta\epsilon_i(t)$ and
\begin{equation}
    \begin{cases}
        \theta_{-1}(t) = \frac{1}{4} \left( \epsilon_1(t) + \epsilon_2(t) + \epsilon_3(t) + \epsilon_4(t) \right)\\
        \theta_{0}(t) = \frac{1}{4} \left( \epsilon_1(t) - \epsilon_2(t) + \epsilon_3(t) - \epsilon_4(t) \right)\\
        \theta_{1}(t) = \frac{1}{4} \left( \epsilon_1(t) + \epsilon_2(t) - \epsilon_3(t) - \epsilon_4(t) \right)\\
        \theta_{2}(t) = \frac{1}{4} \left( \epsilon_1(t) - \epsilon_2(t) - \epsilon_3(t) + \epsilon_4(t) \right).
    \end{cases}
\end{equation}
While for the off-diagonal terms, we have
\begin{equation}
\begin{split}
    H^\text{od} =& \left( V_{12} \ketbra{1}{2} + V_{23} \ketbra{2}{3} + V_{34}\ketbra{3}{4}+ V_{14} \ketbra{1}{4} \right) + h.c.\\
    \xrightarrow{\text{to qubits}}&
    V_{12} \ketbra{00}{01} + V_{23} \ketbra{01}{10} + V_{34} \ketbra{10}{11} + V_{14} \ketbra{00}{11} + h.c.\\
    \xrightarrow{\text{to Pauli}}&
    \theta_{3} \mathbb{I} \otimes \sigma^x +
    \theta_{4} \sigma^z \otimes \sigma^x +
    \theta_{5} \sigma^x \otimes \sigma^x +
    \theta_{6} \sigma^y \otimes \sigma^y.
\end{split}
\end{equation}
where $\epsilon_i(t) = \epsilon_i + \delta\epsilon_i(t)$ and
\begin{equation}
    \begin{cases}
        \theta_{3} = \frac{1}{2} \left( V_{12} + V_{34} \right)\\
        \theta_{4} = \frac{1}{2} \left( V_{12} - V_{34} \right)\\
        \theta_{5} = \frac{1}{2} \left( V_{23} + V_{14} \right)\\
        \theta_{6} = \frac{1}{2} \left( V_{23} - V_{14} \right).
    \end{cases}
\end{equation}

\begin{figure*}
    \centering
    \includegraphics[width=0.75\textwidth]{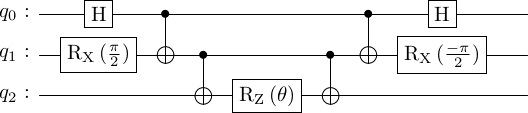}
    \caption{CNOT-staircase decomposition of the unitary operator $U = \exp \left( -i \theta \sigma^x_0 \sigma^y_1 \sigma^z_2 / 2 \right)$.}
    \label{fig:cnot_staircase}
\end{figure*}

The propagator associated with the Hamiltonian in eq. \ref{app_eq:pauli_string_H} can be implemented using first-order Trotter decomposition as
\begin{equation}
    e^{-i H \Delta t} \approx \left( \prod_k e^{-i h_k \Pi_k \Delta t / m} \right)^m,
\end{equation}
and the unitary evolutions of the Pauli strings, $\exp{-i h_k \Pi_k \Delta t / m}$, can be implemented using the well-known CNOT-staircase method \cite{Jaderberg2022,Sawaya2020,Yordanov2020}, an example of which is depicted in figure \ref{fig:cnot_staircase}.
The CNOT staircase can be constructed by following the instructions:
\begin{enumerate}[label=(\alph*)]
    \item Qubits associated with the identity operator in the Pauli string are not involved;
    \item To the other qubits, apply the following gates:
    \begin{equation}
    \begin{cases}
        \mathbb{I} & \text{if }\sigma^z\text{ in the Pauli string}\\
        \text{H} & \text{if }\sigma^x\text{ in the Pauli string}\\
        \text{R}_\text{X}(\pi/2) & \text{if }\sigma^y\text{ in the Pauli string}
    \end{cases}
    \end{equation}
    \item Append cascading CNOT gates by connecting consecutive qubits involved in the Pauli string;
    \item Apply a partial Z-rotation, $\text{R}_\text{Z}(2 h_k \Delta t)$, to the final qubit of the cascade;
    \item Apply the reverse cascade of CNOTs defined in steps (c);
    \item Apply the adjoint gates defined in steps (b).
\end{enumerate}
By employing the CNOT-staircase method, the time evolution of a Pauli string requires $2\tilde n - 2$ CNOT gates, where $\tilde n \le n$ is the number of qubits involved in the Pauli string.
As a last remark, we recall that a partial rotation is defined as $\text{R}_\sigma(\theta) = \exp{-i \theta \sigma / 2}$.
%

\section{Hamiltonians} \label{app:Hamiltonians}
The system used as a case study in Section \ref{subsec:Quantum emulation} and Section \ref{subsec:Effects of memory time on transport efficiency} has a first-exciton manifold Hamiltonian which reads
\begin{equation} \label{eq:Hamiltonian_example}
    H =
    \renewcommand{\arraystretch}{0.75}
    \begin{bmatrix}
      0.442 & 1 & 0 & 1\\
      1 & 0.233 & 1 & 0\\
      0 & 1 & -3.227 & 1\\
      1 & 0 & 1 & 0.356
    \end{bmatrix}.
\end{equation}
This Hamiltonian was previously used in ref. \cite{Gallina2022} for simulating quantum algorithms for computing ENAQT in a Markovian environment.

The system that displays memory-assisted quantum transport in Section \ref{subseq:Memory-assisted quantum transport} has a first-exciton manifold Hamiltonian which reads
\begin{equation} \label{eq:Hamiltonian_MAQT}
    H =
    \renewcommand{\arraystretch}{0.75}
    \begin{bmatrix}
      -1.504 & 1 & 0 & 1\\
      1 & 0.491 &  & 0\\
      0 & 1 & -2.643 & 1\\
      1 & 0 & 1 & 1.436
    \end{bmatrix}
\end{equation}
whose eigenvectors are
\begin{equation} \label{eq:eigenvectors_MAQT}
    W =
    \renewcommand{\arraystretch}{0.75}
    \begin{bmatrix}
      0.3128 & -0.8746 & -0.1707 &  0.3287\\
      -0.3107 & 0.1900 &-0.8614 & 0.3540\\
      0.8625 & 0.4248 &-0.1146 & 0.2502\\
      -0.2486 & 0.1358 & 0.4644 & 0.8391
    \end{bmatrix}.
\end{equation}
%

\section{Measure of trajectories} \label{app:measure}
Here, we comment on the number of measurements needed for the estimation of a population $P_i(t)$ of the open quantum system from a quantum simulation, which is the relevant case for our measure of ENAQT. In this case, the expectation value is bounded between $0$ and $1$ and can be estimated by measuring qubit $i$ along the Z-axis with the physical mapping, or by measuring the probability of the corresponding state with the algorithmic mapping.
From eq. \ref{eq:exact_expectation}, the expectation value of the observable can be written as 
\begin{equation} \label{sup_eq:estimation_P}
    P_i(t) = \lim_{\Delta t \rightarrow 0} \lim_{\Xi \rightarrow \infty} \frac{1}{\Xi} \sum_{\xi=1}^{\Xi} \left( \lim_{K \rightarrow \infty} \frac{1}{K} \sum_{k=1}^{K} b_{k,\xi} \left[ P_i(t=S\Delta t) \right] \right),
\end{equation}
where $K$ is the number of measurements (also called shots) and $b_{k,\xi} \left[ P_i(t=S\Delta t) \right]$ are the measurement outcomes in the form of Bernoulli trials (0 or 1). Because the proposed algorithm relies on computing the open system dynamics as an average over stochastic trajectories, we leverage the averaging procedure to take a single sample for each stochastic realization of the quantum circuit. Indeed, eq. \ref{sup_eq:estimation_P} is equivalent to
\begin{equation}
     P_i(t) = \lim_{\Delta t \rightarrow 0} \lim_{\Xi \rightarrow \infty} \frac{1}{\Xi} \sum_{\xi=1}^{\Xi} b_{\xi} \left[ P_i(t=S\Delta t) \right].
\end{equation}
For any finite discretization of $\Delta t$, the estimation of the observable is then obtained as
\begin{equation}
    \overline{P_i(t=S\Delta t)_\xi} = \lim_{\Xi \rightarrow \infty} \frac{1}{\Xi} \sum_{\xi=1}^{\Xi} b_{\xi} \left[ P_i(t=S\Delta t) \right]
\end{equation}
with an associated error that, in the limit of large $\Xi$, approaches that of the binomial distribution 
\begin{equation}
    \text{error} = \sqrt{\frac{\overline{P_i(t=S\Delta t)_\xi} \left( 1 - \overline{P_i(t=S\Delta t)_\xi} \right)}{\Xi}}.
\end{equation}
%

\section{Scaling of the quantum algorithm} \label{app:scaling}
\begin{figure*}
    \centering
    \includegraphics[width=\textwidth]{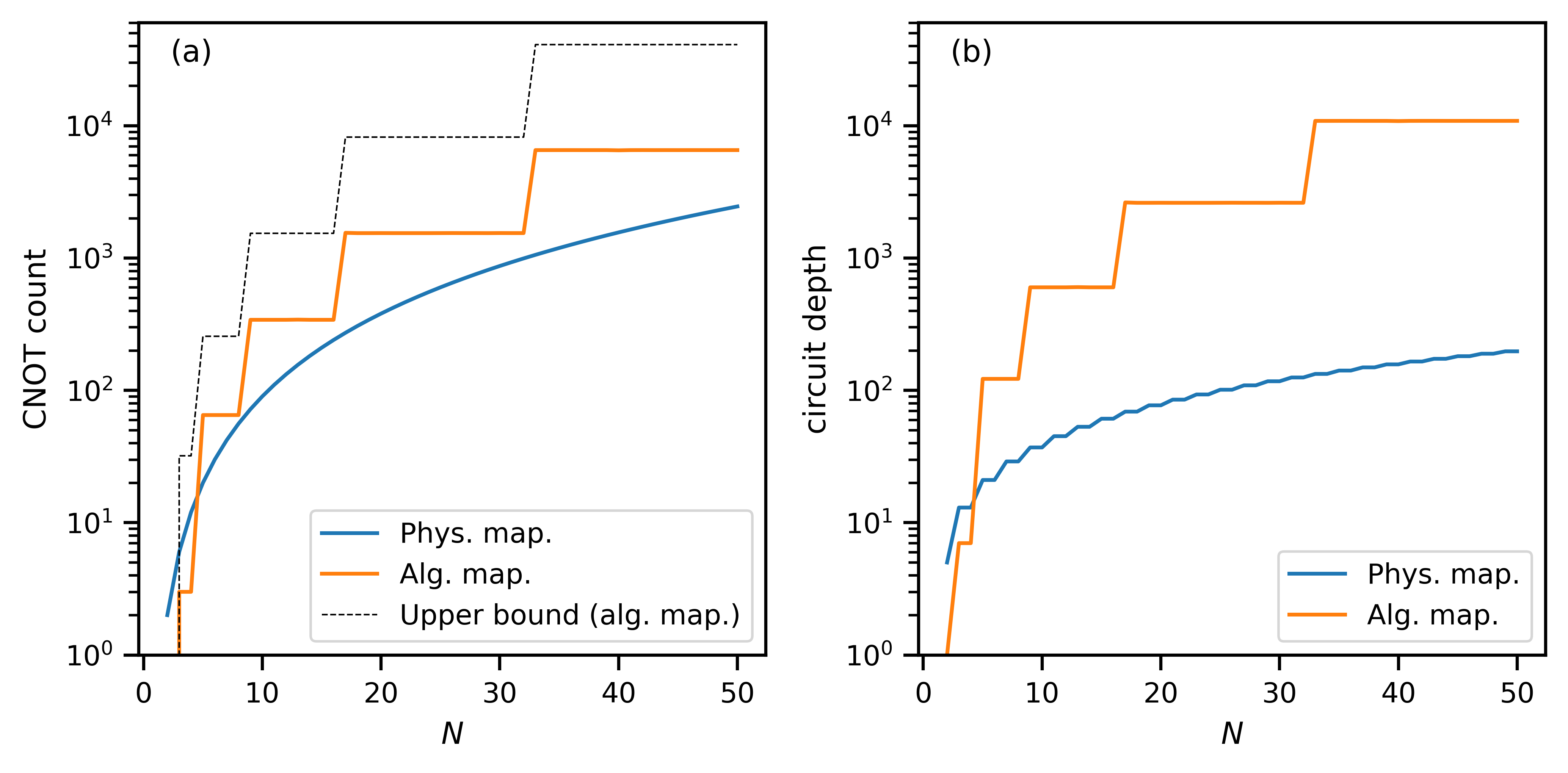}
    \caption{Algorithmic requirements for the implementation of a Trotter step with the physical (blue) and algorithmic (orange) mapping. The CNOT gate count (a) and circuit depth (b) are considered. The upper bound of the CNOT count found for the algorithmic mapping (black line) is also reported.}
    \label{fig:scaling}
\end{figure*}

Here, we analyze the number of CNOT gates needed to execute one Trotter step and its circuit depth, which indicates the minimum number of layers of parallel operations required to obtain the output.
Consequently, this number of operations must be multiplied by the Trotter number $m$ and by the number of time steps $S$ to get the full circuit at time $t = S \Delta t$.
Circuits are built architecture-free, so we do not include any SWAP gate that may be required to ensure communication between non-adjacent qubits in a real device with fixed topology.
As a testbed, we choose a network with long-range interactions between all sites, corresponding to a dense Hamiltonian matrix in the algorithmic mapping with $N^2$ non-zero entries randomly drawn from a normal distribution.
The problem corresponds to the worst-case scenario in terms of required resources.
The scaling obtained for the implementation of circuits composed of CNOT gates and single qubit rotations (U gates) is shown in figure \ref{fig:scaling}. A minimum optimization for the circuit has been used to remove redundant gates (such as consecutive CNOT gates acting on the same qubits).

When using the physical mapping of the system, the results are analytical.
The number of CNOT gates per Trotter step scales proportionally to the number of two-body interactions, that is $N(N - 1)/2$.
In eq. \ref{eq:Trotter_physical}, we employed $\text{R}_\text{XX}$ and $\text{R}_\text{YY}$ gates to take into account interactions between pairs of chromophores.
One pair of $\text{R}_\text{XX}$ and $\text{R}_\text{YY}$ gates acting on the same pair of qubits can be optimally decomposed into a sequence of universal gates that include only 2 CNOT gates.
Therefore, $N(N-1)$ CNOT gates are required for the implementation (blue line in figure \ref{fig:scaling}a).
Remarkably, the circuit depth required for a physical mapping scales with $\mathcal{O}(N)$, as multiple operations can be performed in parallel. In fact, when $N$ is even, the circuit can be rearranged to form $N-1$ layers of operations between $N/2$ pairs of unique qubits that can be executed simultaneously.
On the other hand, when $N$ is odd, the circuit can be arranged with $N$ layers involving operations between $(N-1)/2$ unique qubits.

For the algorithmic mapping, obtaining an exact scaling of the number of CNOT gates is more complex.
Examining eq. \ref{eq:Hamiltonian_to_Pauli}, which decomposes the system Hamiltonian into Pauli strings, reveals that each qubit can contribute to the string with four potential Pauli operators ($\mathbb{I}$, $\sigma^x$, $\sigma^y$ or $\sigma^z$).
If we take $n = {\lceil \log_2 N \rceil}$ as the number of qubits used for the implementation, then the Hamiltonian is composed of $4^n$ Pauli strings (which is $N^2$ in the case $N=2^n$).
Now we assume that each string involves all the qubits of the quantum register, which is strictly true only for strings that do not contain the identity operator.
Then, using the CNOT staircase method, we need $4^{n+1/2}(n-1)$ CNOT gates for producing the quantum circuit for the complete Hamiltonian propagator (black line in figure \ref{fig:scaling}a).
Since this represents the limit case, as can be seen by looking at the CNOT count coming from an implementation of the circuit (orange line in figure \ref{fig:scaling}a), we can confidently state that the upper bound is $\mathcal{O}\left( 4^n n \right) = \mathcal{O}\left( N^2 \log_2 N \right)$.
As we could expect, from figure \ref{fig:scaling}a, we can also notice that the number of required CNOT gates is constant for networks composed of the same number of chromophores.
Further approaches can be used to reduce the gate demand.
For example, based on the network topology, different binary mappings can lead to optimized counts (for example, Gray code can be beneficial for linear chains or cyclic mappings \cite{Sawaya2020}).
Finally, we comment on the circuit depth (orange line in figure \ref{fig:scaling}b).
Unlike physical mapping, where an optimal arrangement of two-qubit interactions can reduce the circuit depth, algorithmic mapping involves operations that act on a significant portion of the quantum register.
Therefore, most of the gates are executed in series and the number of operations that can be performed in parallel is limited.

To conclude, despite the difference between linear and logarithmic qubit requirements, physical mapping exhibits some advantages over algorithmic mapping in terms of CNOT gate usage and circuit depth.

%
\printbibliography
\end{document}